\documentclass[onecolumn,showpacs,preprintnumbers,amsmath,amssymb,superscriptaddress]{revtex4}

\usepackage{amssymb}

\usepackage{txfonts}

\usepackage{bbold}

\usepackage[pdftex,colorlinks=true,bookmarks=false,citecolor=blue,urlcolor=blue]{hyperref} 

\usepackage{graphicx}
\usepackage{dcolumn}
\usepackage{bm}

\usepackage{CJK}

\usepackage{natbib}
\begin{document}

\begin{CJK*}{GB}{gbsn}

{Subject Area: \u{C}erenkov radiation; Metamaterials; Light Source; Electromagnetic optics\\
Correspondence and requests for materials should be addressed to L. Y. (liuy0074@e.ntu.edu.sg)}

\title{Motion-induced radiation from electrons moving in Maxwell's fish-eye}

\author{Liu Yangjie(Áõãó½Ü)}

\email[Electronic address of corresponding author:] 
{liuy0074@e.ntu.edu.sg}
\affiliation{Computational Nano-Electronics Laboratory, Division of Micro-Electronics, School of Electrical and Electronic Engineering, 50 Nanyang Avenue, Nanyang Technological University, Singapore 639798}

\author{L. K. Ang}
\email{ricky\_ang@sutd.edu.sg}

\affiliation{Singapore University of Technology and Design, 20 Dover Drive, Singapore 138682}

\date{Started 11 June, 2012; revised 25 Sept 2013, compiled \today}

\begin{abstract}

\noindent In \u{C}erenkov radiation and transition radiation, evanescent wave from motion of charged particles transfers into radiation coherently. However, such dissipative motion-induced radiations require particles to move faster than light in medium or to encounter velocity transition to pump energy. Inspired by a method to detect cloak by observing radiation of a fast-moving electron bunch going through it by Zhang {\itshape et al.}, we study the generation of electron-induced radiation from electrons' interaction with Maxwell's fish-eye sphere. Our calculation shows that the radiation is due to a combination of \u{C}erenkov radiation and transition radiation, which may pave the way to investigate new schemes of transferring evanescent wave to radiation.










\end{abstract}

\maketitle

\end{CJK*}


{Transfer from evanescent wave into propagating wave is an interesting optical problem~\cite{Renger2010}, which is still relatively unexplored perhaps because of the distinct feature of evanescent wave. 
Evanescent wave appears along with moving electrons, and it does not carry power unless electrons are moving faster than light in a medium, which is known as \u{C}erenkov radiation.  
\u{C}erenkov radiation occurs when fast motion of electrons leaves their accompanying evanescent fields lagged behind in space, which will coherently accumulate into energetic radiation, similar to the sonic boom created by an object travelling faster than sound. 
A different type of radiation called transition radiation is able to bypass the requirement of fast velocity for electrons, which occurs as long as electrons move through an inhomogeneous media, such as a boundary between two different materials. 
It is worth mentioning here that \u{C}erenkov radiation from photonic crystals~\cite{Luo2003} is a special case because in photonic crystal, the loss of translational invariance will allow emission of radiation without any threshold velocity of electrons, which has intrinsically coupled with transition radiation. 

A third type of radiation also transforming evanescent wave into radiation from electrons' motion is known as diffraction radiation, for which electrons generate radiation in passage near a structured surface without penetrating it. 
This mechanism has been shown recently in a metamaterial-based structure, which is driven by a free electron beam bunch~\cite{Adamo2009}. 
In this light-well source, free electrons interact with periodically-alternating environment as they travel through tunnels inside, and it is proved theoretically that the radiation is due to the diffraction radiation ~\cite{Liu_SG2011}, which shows a continuum of radiation over a broad angular range. 
This transfer from kinetic energy of an electron bunch to radiation also bears resemblance to the transition radiation because electrons also interact with inhomogeneous dielectric parameters $\epsilon(\bar r)$ during their passage. 
This proof-of-concept not only opens up new type of electron-induced radiation sources, but also inspires one to ponder whether other inhomogeneous structure can induce similar radiation phenomenon. 


The recently-developed field of transformation optics~\cite{Pendry2006, Leonhardt2010book} provides the capability to manipulate light flow almost arbitrarily, for instance to hide objects inside a cloak device made of inhomogeneous and anisotropic permittivity and permeability. The invariant form of Maxwell's equations, under various coordinates (including even non-Euclidean systems), allows light to be shaped by curved electromagnetic space in order to design novel devices for various applications. 
From the perspective of an electron, it does not bend the trajectory similarly to light and instead experience a curved space to induce radiation~\cite{Zhang2009}. 
Therefore such a curved electromagnetic space becomes an energy-pumping candidate to give rise to electron's motion-induced radiation in higher frequency with assist of nanofabrication technology~\cite{Valentine2009,Zentgraf2011}. 
In particular, Zhang {\itshape et al.} invented a method to detect cloak by observing radiation of a fast-moving electron bunch going through the cloak device~\cite{Zhang2009}. 
In this paper, we are interested to study the electron's motion-induced radiation for electron bunches moving through a different structure known as Maxwell's fish-eye, which in principle will provide unlimited resolution as a perfect imaging lens ~\cite{Leonhardt2009a}.
From our calculation, we will show that the the radiation is a combination of both \u{C}erenkov radiation and transition radiation. 

The rest of this paper is structured as follows. 
For Results section, in Subsection A the basic equations of electromagnetic fields due to moving charges are posed and the analytic solutions to electromagnetic fields are derived by dyadic Green's function.  In Subsection B main calculation results are shown including electric fields on xz plane (where electron's trajectory is confined) and field radiation patterns at a frequency of 300 ${\rm THz}$. 
In Discussion Section, we analyze and discuss temporal evolution of electron's radiation from Maxwell's fisheye sphere and then summarize this report with some outlook to the potential application of this work. The formulation of dyadic Green's function method is reported in Method Section and Supplementary information. }

\section*{Results}

\subsection*{A. Question of radiation from the Maxwell's fisheye sphere}

In this section we shall pose our problem and try to give some qualitative prediction from the perspective of transformed uniform space. The Maxwell's fisheye lens of refractive index profile
\begin{eqnarray}\label{}
n(r)&=&\frac{2}{1+(\frac{r}{R_1})^2} , r \in [0,\infty),
\end{eqnarray}
invented by Maxwell himself, focuses light-rays emitted from a source point to its image point antipodally. In this manuscript, we investigate two cases of Maxwell fisheye profile: one is a nonmagnetic Maxwell fisheye sphere, whose permittivity and permeability functions are defined as
\begin{eqnarray}
\label{epsr1}
\epsilon_r(r)&=&n^2(r) , r \in [0, R_1]\\
\label{mur1}
\epsilon_r(r)&=&1, r\in (R_1,\infty)\\
\mu_r(r)&=&1. r \in (0,\infty)
\end{eqnarray}
For outer space of the Maxwell's fisheye sphere, simply vacuum permittivity is adopted and the whole space we investigate is un-magnetic in order to facilitate possible future experimental work. The other is the impedance-matched full Maxwell fisheye, $\epsilon=\mu=n(0<r<\infty)$. As the surface of a sphere is a curved space(with constant curvature), the fish eye profile $n(r)$ can be mapped onto the surface of a hypersphere(4D manifold) in virtual uniform space from Lunburg's visualization~\cite{Leonhardt2009b}. For simplicity, suppose we generate a pulse of electron beam $\bar J(\bar r)$ moving along a straight line in the $z$-direction inside a Maxwell's fisheye sphere,
\begin{eqnarray}\label{Jrt}
{\bar J}({\bar r},t)={\hat z}qv\delta(x-x_0)\delta(y-y_0)\frac{1}{\sigma\sqrt{2\pi}}\exp\Big[{-\frac{(z-vt)^2}{2\sigma^2}}\Big].
\end{eqnarray}
Our goal is to obtain the solution to inhomogeneous wave equations of electromagnetic fields from dyadic Green's function method analytically~\cite{Tai1990,Tai1993,Kong_EWT}. In this manuscript, we take $q=1000e,R_1=2{\rm \mu m}, x_0=1{\rm \mu m}, y_0=0, \sigma=50{\rm nm}, v=0.9c$(we use 1000 electrons to represent an electron bunch, $e$ indicates elementary charge and $c$ light velocity in vacuum). The formulation of the dyadic Green function to solve Maxwell's equations shall be demonstrated in Methods section. Only dyadic Green function in the source region(the nonmagnetic Maxwell's fisheye sphere) is given here: 
\begin{eqnarray}\label{Geo}
\bar{\bar{G}}_{eo}=-\frac{1}{k^2}\hat{r}\hat{r}\delta(\bar{r}-\bar{r}')+\frac{ik}{4\pi}\sum_{m,n}C_{mn}\\\nonumber
\begin{cases}
\bar{M}^{(m1)}(k) \bar{M}'^{(m)}(k) + \bar{N}^{(e1)}(k)\bar{N}'^{(e)}(k), & r>r',\\
\bar{M}^{(m)}(k) \bar{M}'^{(m1)}(k) + \bar{N}^{(e)}(k)\bar{N}'^{(e1)}(k),& r<r',\\
\end{cases}
\end{eqnarray}
in which $k$ is the wavevector, $\bar{M}^{(m1)},\bar{M}^{(m)},\bar{N}^{(e1)},\bar{N}^{(e)}$ are the eigen-wave vectors in the Maxwell's fisheye sphere, $C_{mn}$ the coefficient for the infinite series. Thus electric field in in the Maxwell's fisheye sphere can be written as 
\begin{equation}\label{Efieldformula}
\bar{E}(\bar{r}) = i\omega\mu_0\iiint\bar{\bar{G}}_{eo}(\bar{r}, \bar{r'})\cdot\bar {J}(\bar {r'})\mathrm d V'. 
\end{equation}
The Green function for electromagnetic fields in the impedance-matched full Maxwell fisheye has been solved recently\cite{Leonhardt2010b, Leonhardt2011}. The Green tensor for electric field should be, 
\begin{eqnarray}\label{GtensorUlf}
G(\bar r,\bar r', k)=\frac{\nabla\times n(r,r')\nabla\otimes\nabla' D(r,r')\times \overleftarrow \nabla'}{n(r)n(r')k^2}-\frac{\delta(\bar r-\bar r') \mathbb {1}  }{n(r)k^2},
\end{eqnarray}
in which $r$ and $r_0$ are taken in the relative unit of inner sphere radius $R_1$ and light speed $c=1$ to avoid unnecessary factors(See  Supplementary information III for its explicit form), however one requires to convert the unit after obtaining the physical quantities. Thus electric field in the impedance-matched Maxwell's fisheye can be written the same as equation ~\eqref{Efieldformula}.

\begin{figure}[htbp]
  \centering
\includegraphics[width=1\columnwidth]{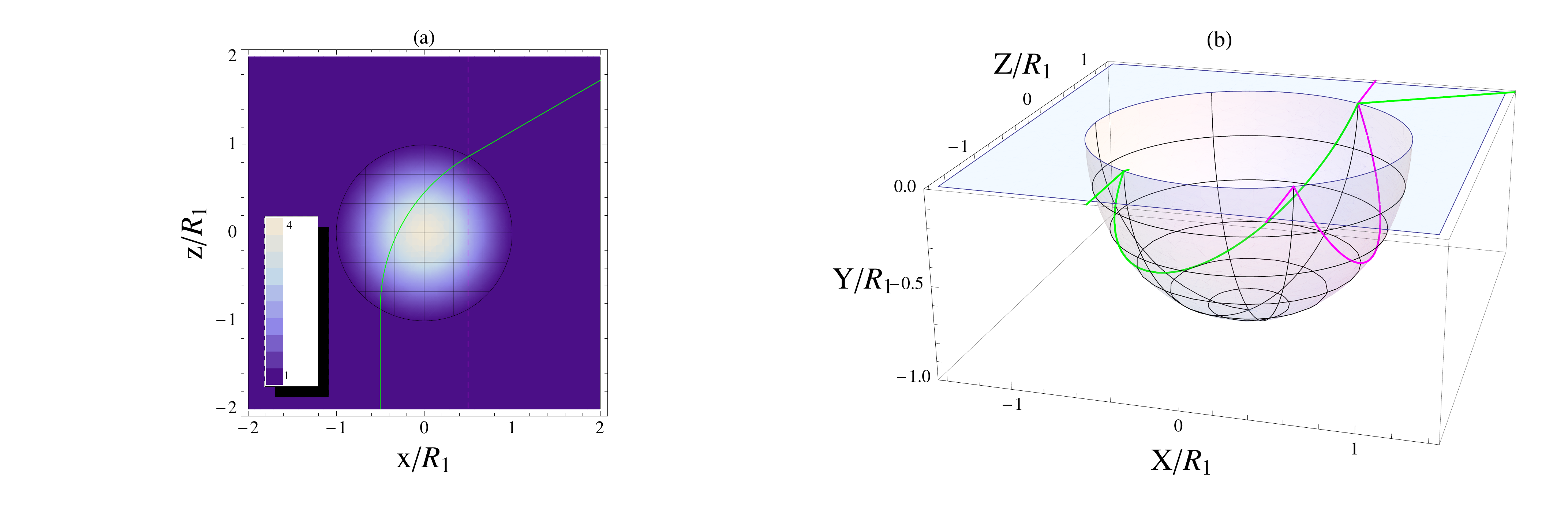}

\caption
{\label{fig:projection}
(Color online) Physical space (upper panel (a)) can be transformed into virtual space lower panel (b) which is curved space under inverse stereographic projection. In both panels green lines indicate the light trajectory while the magenta the electron's trajectory. 
Panel (a) is the physical space of xz plane with permittivity distribution $\epsilon_r(r)$ marked in blue-color, which can be divided into two parts demarcated by the black circle: inner part $0\leq r\leq R_1$ is Maxwell's fisheye and outer part $r>R_1$ is uniform space($\epsilon_r(r)=1$). In colorbar, brighter color indicates the higher value of permittivity. The magenta line indicates the trajectory of electron bunch along straight line $x={R_1}/{2}, y=0,z=0$.  
Panel (b) is the virtual uniform space, composed of two parts corresponding to physical space. One is invariant planar space marked in light blue, $Y=0, X^2+Z^2\geq R_1^2$; the other lower half spherical surface in uniform space in pink (mapped from Maxwell's fisheye sphere on xz plane zone, $Y=0, X^2+Z^2< R_1^2$ shown in panel (a), according to inverse stereographic projection). Note: green and magenta curve intersects at point $x={1}/{2}, z={\sqrt 3}/{2}, y=0$. This is solely coincidence due to the parameters we choose. 
}
\end{figure}

Before any further calculation, it is suggestive to analyze how this curved geometry manipulates different dynamics of light and electron both in physical space and virtual uniform space. In panel(a) of Fig.~\ref{fig:projection}, permittivity distribution on xz plane is demonstrated in bluecolor. According to Hamilton's  equations $\dot{\bar r}=\partial \omega/\partial \bar k$ and $\dot{\bar k}=-\partial \omega/\partial \bar r$,  in which $\bar k$ indicates momentum, $\omega$ the Hamiltonian~\cite{Ochiai2008, Leonhardt2010book}, the light trajectory in xz plane can be traced to follow an arc of a circle within Maxwell's fisheye in green curve, in physical space in panel(a) of Fig.~\ref{fig:projection}; whereas motion of charge does not bend its trajectory at all in dashed magenta line according to our predetermined current density function~\eqref{Jrt}. Now we transform physical space to virtual uniform space to contrast the dynamics of light and charge. Since it is generally difficult to plot surface of a 4-dimensional hypersphere from our human perspective, we reduce this to a simpler counterpart without losing its curved property: surface of a 3-dimensional sphere(panel(b) of Fig.~\ref{fig:projection}) and limit our view to the cross section in $xz$ plane. Mapped to uniform space, light follows geodesics(the green curve) on the lower half (corresponding to interior Maxwell's fisheye)spherical surface. However, the electrons deviate their trajectory from geodesics of spherical surface in uniform space if we transform their trajectory to uniform space(X, Z, Y) from physical space (x,z), according to inverse stereographic projection~\cite{Leonhardt2010book}(we change sequence of Y and Z to accommodate xz plane in physical space),
\begin{eqnarray}
X=\frac{2x}{1+x^2+z^2}, \\
Z=\frac{2z}{1+x^2+z^2},\\
Y=\frac{x^2+z^2-1}{x^2+z^2+1}.
\end{eqnarray}
In uniform space, electrons propagate along a curve on lower half spherical surface, shown in dashed magenta circle in Fig.~\ref{fig:projection}(b). From the bent trajectory of charge in virtual uniform space, we can explain why electrons radiate in our question. In this stretched uniform space, electrons actually experience a curved path on lower spherical sphere. This predicts that the electrons will generate a synchrotron radiation within Maxwell's fisheye sphere, in agreement with our results below. 

Notice this geometric picture of electron moving on a sphere applies \emph {exactly only} under the condition of \emph {impedance-match}, $\epsilon=\mu=n(0<r<\infty)$. While the \emph {nonmagnetic} case, $\epsilon=n^2(r),\mu=1$, within Maxwell's fisheye, is derived under the eikonal approximation which allows one to vary permittivity and permeability so that refractive index maintains unchanged~\cite{Landy2012}, however at the price of varying the wavefront of electromagnetic fields to a certain extent(this 3D case could not guarantee a trick to keep the polarization under eikonal approximation instead of the 2D case in Ref.~\cite{Leonhardt2009a}). Therefore the geometric picture in \ref{fig:projection}(b) is only an \emph{approximation} under this nonmagnetic rescaling in equations~\eqref{epsr1} and \eqref{mur1}. However, we shall demonstrate in the next subsection, that the two cases both induce a combination of \u{C}erenkov radiation and transition radiation, and thus our approximation remains valid with regard to the perspective of radiation.

\subsection*{B. Main results}

\begin{figure}[htbp]
  \centering
	\includegraphics[width=1\columnwidth]{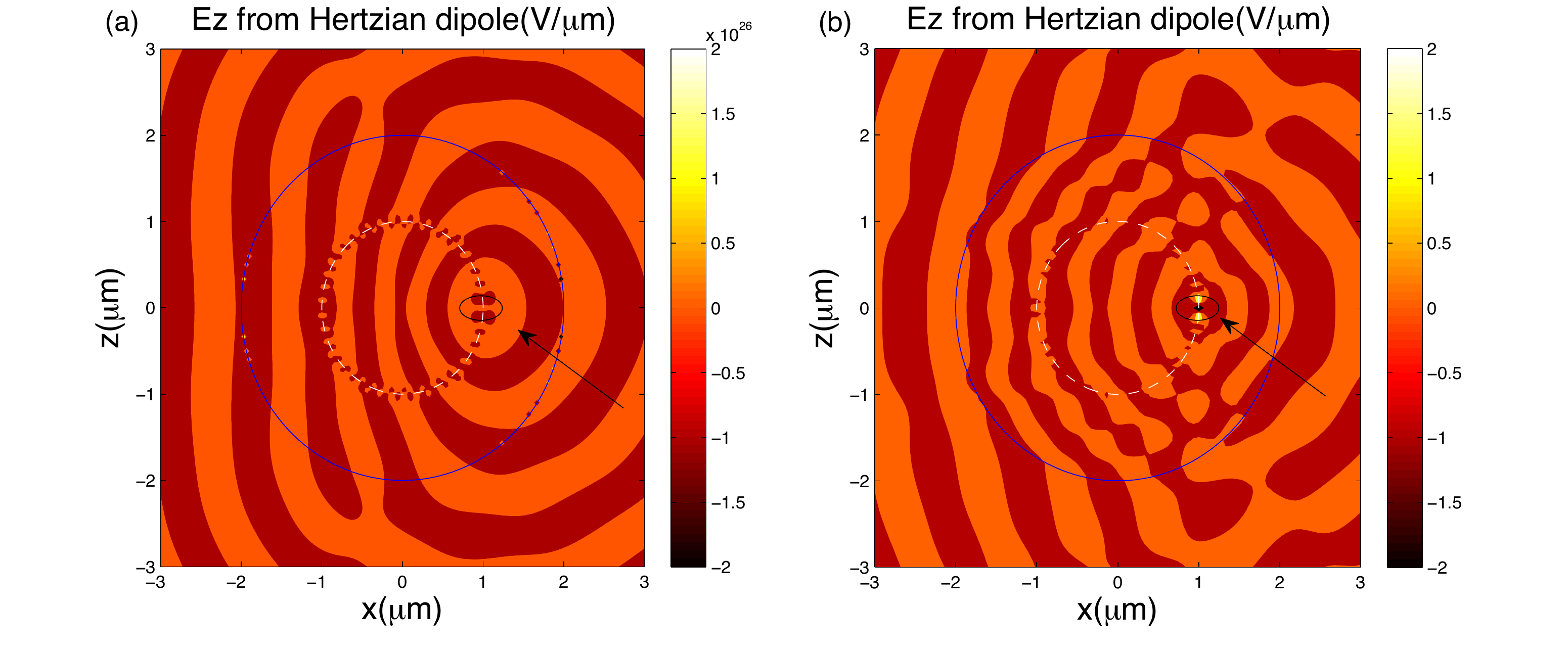} 
\caption
{\label{fig:E3dip}
(Color online) Electric fields in z direction, $E_z(x,z)$ on xz plane from Hertzian dipole at point $x=R_1/2=1\mu m, y=0, z=0$(marked as a small black circle) for (a) nonmagnetic Maxwell's fisheye sphere(c.f. also FigS1(a) for impedance matched case) and (b) uniform sphere ($\epsilon_r(r<R_1=4$). The huge-value points along blue circle in (a) are thought to be numerical defect of hypergeometric function. The dashed lines in white in (a-b) indicate where some discontinuity at circle $r=r'$ is due to piecewise form of dyadic Green function $\bar {\bar {G}}_{eo}$ itself in equation~\eqref{Geo}.  
}
\end{figure}

First to demonstrate the unique electromagnetic geometry of Maxwell's fisheye structure, we position an infinitesimal Hertzian electric dipole $\bar J(\bar r')=-\hat {z} i\omega q l \delta(\bar r'-\bar r_0)$($l$ the dipole length)~\cite{Kong_EWT} at the middle point $\bar r_0$ of the current line, $x=R_1/2=1\mu m, y=0, z=0$, indicated by a black arrow in Fig.~\ref{fig:E3dip}.  In Fig.~\ref{fig:E3dip}, z component of electric field on plane $y=0$ is plotted for single frequency $300{\rm THz}$($\lambda=1\mu m$). Inside Maxwell's fisheye sphere in Fig.~\ref{fig:E3dip}(a), light generates from dipole point and follows a curved path which becomes denser in inner part than in outer(while the counterpart for impedance-matched case is plotted in Fig. S1(a) of Supplementary information III, a similar pattern to Fig.~\ref{fig:E3dip}(a)). This can be explained by the fact that permittivity (or equivalently refractive index) gradually reaches higher in inner part than in outer(also c.f. FigS1(a)) to observe similar phenomenon for impedance-matched Maxwell's fisheye). The dashed line in white indicates the discontinuity at circle $r=r'$, which is due to the piecewise form of dyadic Green function $\bar {\bar {G}}_{eo}$ itself in equation~\eqref{Geo}. Outside the Fisheye, electric field distribution in Fig.~\ref{fig:E3dip}(a) becomes curvilinear smooth contours, different from scattered contours in (b)(permittivity $\epsilon_r=4$ is uniformly distributed inside the blue circle for a contrast). The smooth contours which squeeze inside fisheye implies that the profile of Maxwell's fisheye could in principle be used to design light absorber. 

\begin{figure}[htbp]
  \centering
	\includegraphics[width=1.05\columnwidth]{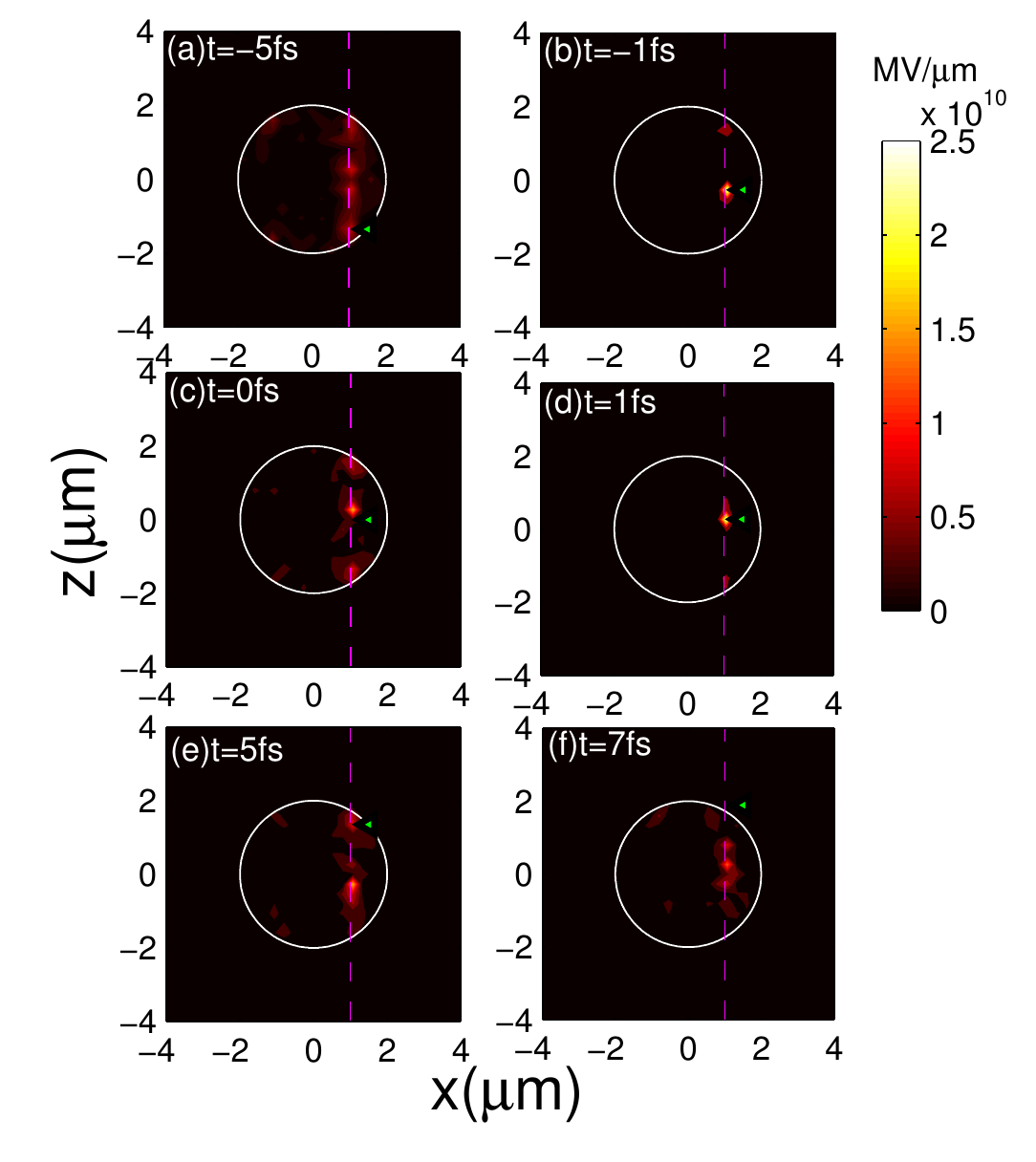}
\caption
{\label{fig:panel}
(Color online) Six snapshots of temporal electric field magnitudes$\vert E(y=0, x,z, t)\vert$ on xz plane for motion-induced radiation from Maxwell's fisheye sphere(nonmagnetic case). Dashed pink lines indicate the trajectory of electrons. Green left arrows indicate the positions of electrons, which are moved by distance of $0.5\mu m $ to the right of their actual positions to avoid shading. We define our time scale coordinate so that electrons pass middle point $x=1\mu m, y=z=0$ at time $t=0$. Parameters: $R_1=2\mu m, q=1000e, v=0.9c$. 
}
\end{figure}

\begin{figure}[htbp]
  \centering
	\includegraphics[width=1.05\columnwidth]{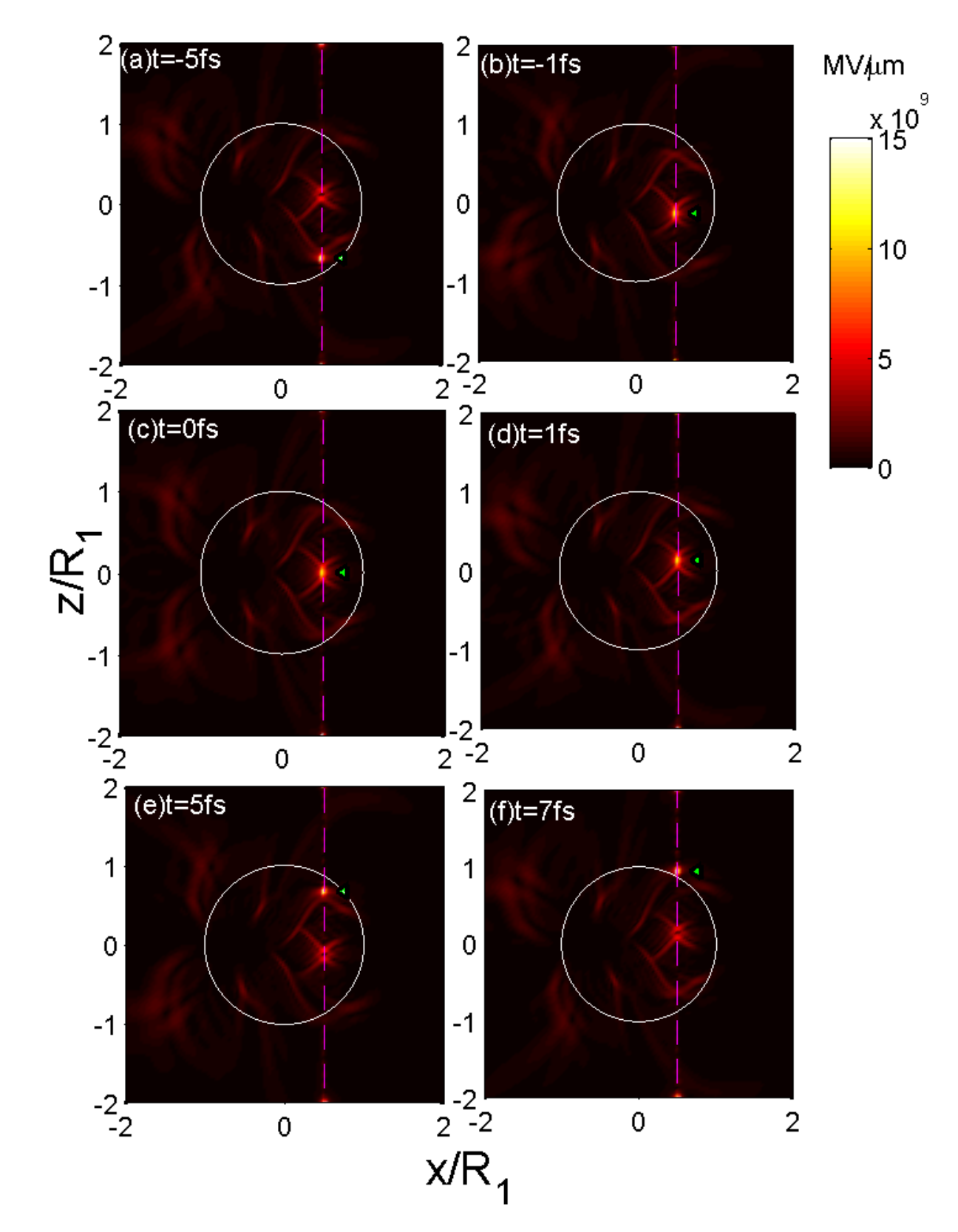}
\caption
{\label{fig:panelnew}
(Color online) Six snapshots of temporal electric field magnitudes$\vert E(y=0, x,z, t)\vert$ on xz plane for motion-induced radiation from Maxwell's fisheye sphere(impedance matched case). Also see the gif file in Supplementary information for transient effect. All the parameters and figure configuration is the same as Fig. ~\ref{fig:panel}. 
}
\end{figure}

Second, we present the magnitude of electric field on xz plane with time. In Figs.~\ref{fig:panel}(nonmagnetic case) and \ref{fig:panelnew}(impedance-matched case) respectively, we present 6 snapshots of magnitudes of the electric field resulting from electron's motion into Maxwell's fisheye sphere. Green left arrows indicate the position of moving electrons, which are moved by distance of $0.5\mu m $ to the right of their actual positions to avoid shading radiation. We define our time scale coordinate so that electrons pass the middle point $x=1\mu m, y=z=0$ at time $t=0$.

\begin{figure}[htbp]
  \centering
	\includegraphics[width=1.\columnwidth]{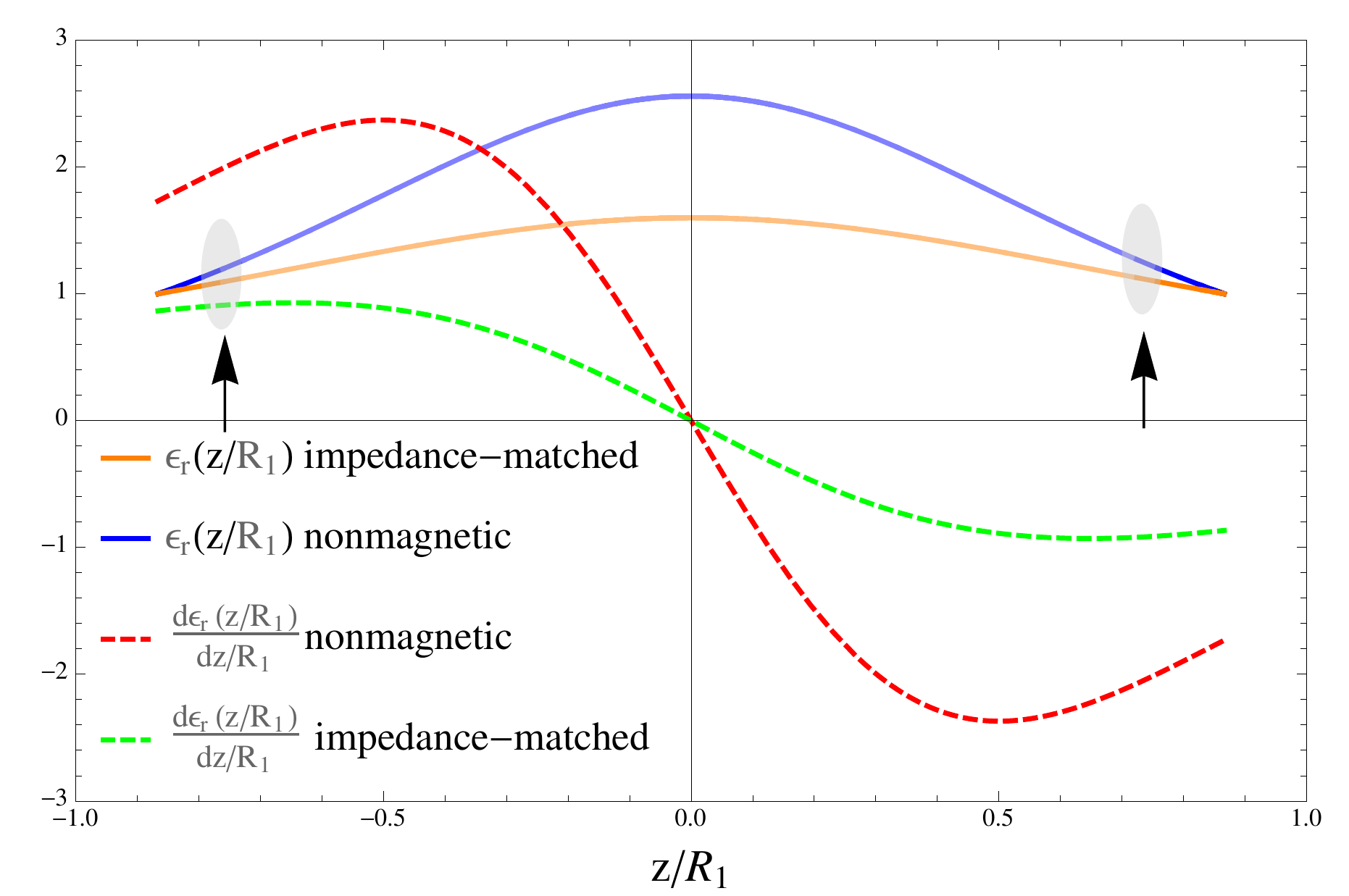}
\caption
{\label{fig:epsz}
(Color online) For Maxwell fisheye in nonmagnetic and impedance-matched cases, permittivity function $\epsilon_r(z/R_1)$and its derivative function ${\rm d}  \epsilon_r(z/R_1)/({\rm d}z/R_1)$ of z coordinate along charge trajectory($x=1\mu m, y=0, -\sqrt{3}/2<z<\sqrt{3}/2$) investigated.  The two arrows indicate the two points where \u{C}erenkov radiation starts and stops, i.e. $n(r)=c/v$.  The interval where \u{C}erenkov radiation lasts is colored lighter in permittivity function lines. 
}
\end{figure}

We first analyze nonmagnetic Maxwell fisheye sphere case in Fig.~\ref{fig:panel}. Electrons generate transition radiation as they enter the Maxwell's fisheye sphere(in Fig.~\ref{fig:panel}(a), only the radiation accompanying electrons are real radiation and other spots in front of them are considered as evanescent waves) because they see different medium upon boundary of fisheye. During the whole process when electrons traverse the inner Maxwell's fisheye, because permittivity varies from low to high and low again along the electron path, transition radiation is generated continuously but not uniformly in time due to inhomogeneous permittivity experienced. We observe stronger radiation mainly from the middle part ((b-d) in Fig.~\ref{fig:panel}) of the trajectory within fisheye. In physical space, permittivity distribution along the electron path in the fisheye informs us the maximum permittivity gradient lies near fisheye's boundary(c.f. red curve in Fig.~\ref{fig:epsz}), which would suggest stronger transition radiation near the boundary instead. However, this disagrees with our calculation results, which can be explained from the perspective of \u{C}erenkov radiation. In dominant part of electron's path inside the inner sphere, permittivity is greater than $({c}/{v})^2(=1.23)$(c.f. light blue curve in Fig.~\ref{fig:epsz}), therefore \u{C}erenkov radiation contributes its power and makes the radiation stronger especially in middle part with higher permittivity. This higher permittivity in middle part could also explain why radiation pulse slows down compared with the electrons during time $t=0\sim1{\rm fs}$ since light speed slows in high permittivity region. As electrons move out of the fisheye, transition radiation bounces back leftward(Fig.~\ref{fig:panel}(e-f)). And there is, however radiation leftover inside fisheye, although electrons are then absent. 

The analyse of impedance-matched full Maxwell fisheye in Fig.~\ref{fig:panelnew} is similar but not identical.  Stronger radiation is also mainly observed in the middle part(panels(b-d) of Fig.~\ref{fig:panelnew}), because \u{C}erenkov radiation(c.f. light orange curve in Fig.~\ref{fig:epsz}) also overweighs and make the whole radiation stronger in the middle part by its higher refractive index(notice that in impedance-matched case, refractive index $n(r)=\epsilon_r(r)$). Transition radiation also occurs since the refractive index is continuously varying along the electrons' path in physical space. However, a \emph{unique} feature of the radiation from impedance-matched Maxwell fisheye is, the electric field accompanying electrons (we call it Electron's Spot(ES) for shortness) carries front and back lobes generally, which switch from the front to the back with time (c.f. (a), (c) and (e) in Fig.~\ref{fig:panelnew}).  In front of or behind field spots , there are other weaker field spots which resembles in shape with ES, except containing a hole in its interior. This type of weak spot keeps rotating away to the left side of electrons' trajectory until it rotates back to align with the electrons' position, when the electrons fill the hole of the weak spot to emanate real radiation from their motion(c.f. gif file in Supplementary information). No scattering light is observed in this case when the electrons move out the the boundary of $R_1$ because impedance-match condition remains valid along this spherical boundary. 

We may understand the transition radiation in impedance-matched Maxwell fisheye from perspective of virtual space. In virtual space in panel(b) of Fig.~\ref{fig:projection}, electrons move along the magenta circle on spherical surface, which is equivalent to a synchrotron radiation from accelerated electrons in uniform medium.  This acceleration includes both bending trajectory and varying velocity. Since electrons move at a fixed speed in physical space, they move at varying speed along a curved path in virtual space. The radiation thus is a synchrotron radiation nonuniform in time.

\begin{figure}[h]
  \centering
	\includegraphics[width=1.1\columnwidth]{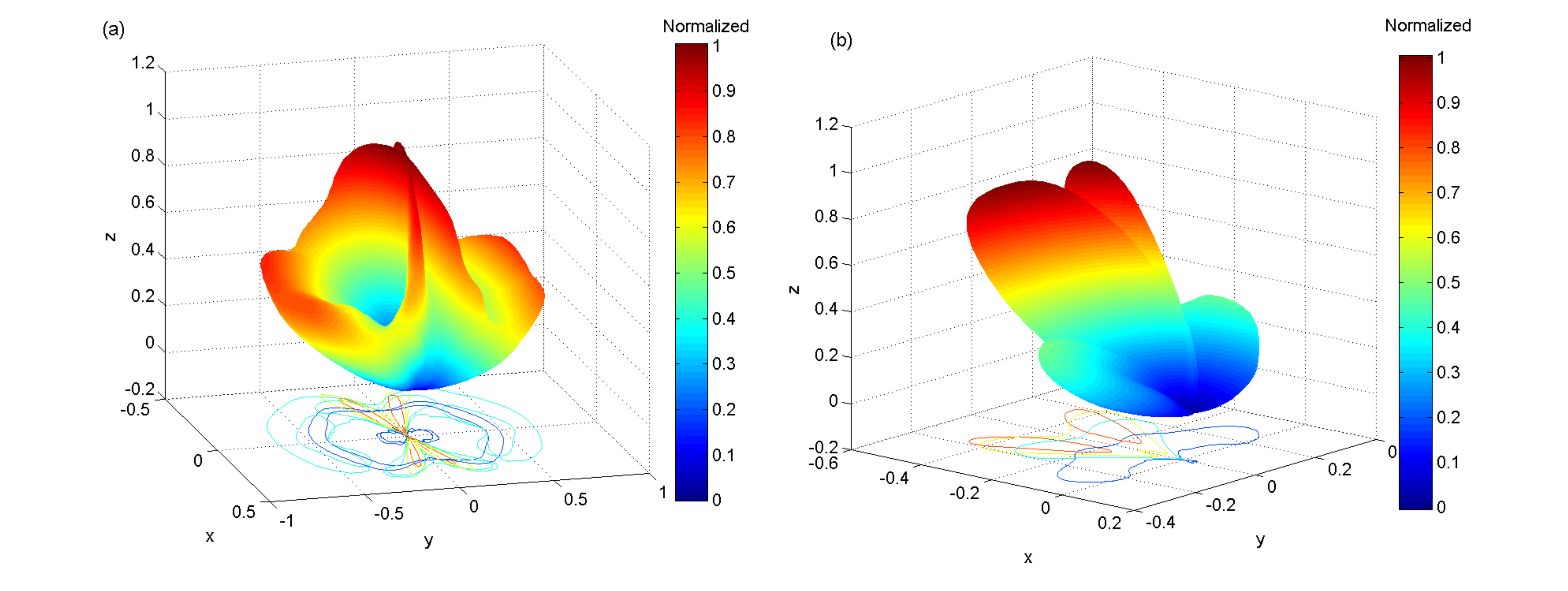}
\caption
{\label{fig:radpat}
(Color online) (a) Near field and (b) far field radiation patterns at $300 {\rm THz}$ from moving charged particle through nonmagnetic Maxwell's fisheye sphere. Enclosing sphere of calculation is double size of fisheye radius $2R_1=4{\rm \mu m}$ and $20R_1=40{\rm \mu m}$. Contours of radiation pattern on xy plane are projected onto ground plane to show asymmetry in x coordinates due to the broken symmetry of injection position.  
}
\end{figure}

\begin{figure}[h]
  \centering
	\includegraphics[width=1.1\columnwidth]{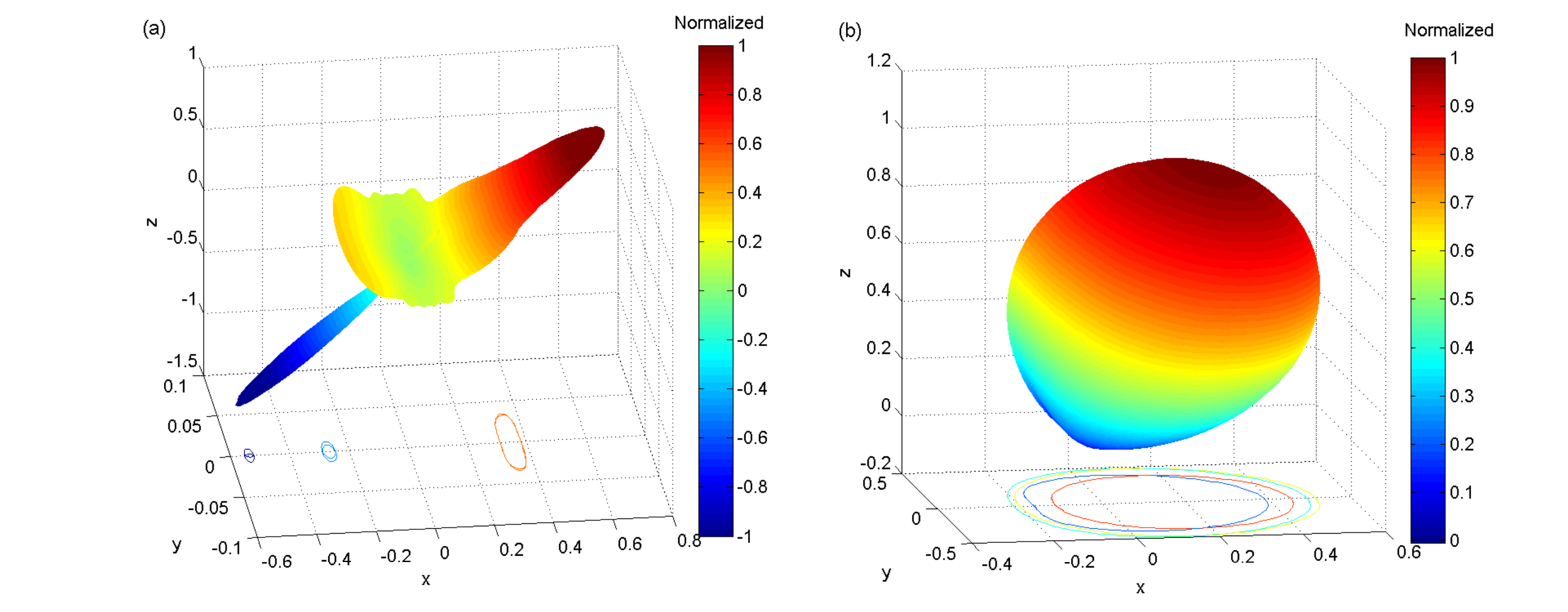}
\caption
{\label{fig:radpat2}
(Color online) (a) Near field and (b) far field radiation patterns at $300 {\rm THz}$ from moving charged particle through the full Maxwell's fisheye for impedance-matched case. Enclosing sphere of calculation is double size of fisheye radius $R_1=2{\rm \mu m}$ and $20R_1=40{\rm \mu m}$. Other configurations of the figure are the same as Fig.~\ref{fig:radpat}. }
\end{figure}

Radiation patterns of near and far field corresponding to a single frequency $300{\rm THz}$ of the whole frequency band, are demonstrated for both sets of the parameters investigated. For nonmagnetic case in Fig.~\ref{fig:radpat}, the two side lobes spread symmetrically in y coordinate and the two radiation patterns are both pointing outwards. While for impedance-matched case in Fig.~\ref{fig:radpat2}, the patterns are distinctly different----the near field one carries a radiation power pointing inwards(minus valued) along the minus-x direction and less side lobes are observed. This negative power flow of radiation can be explained by the rotational motion of the weak field spot with a hole by indicating that more power of radiation flows inward than outward at the minus x side of the boundary $r=R_1$. All four radiation patterns demonstrate that the main lobe of radiation is pointed toward motion direction of electrons, z direction and maintains a symmetrical feature along y coordinate as well as a biased broken symmetry along x coordinate. We attribute this bias of pattern to broken symmetry from biased position of injected electron bunch, given by equation~\eqref{Jrt}.



\section*{Discussion}
From Figs.~\ref{fig:E3dip} and S1(a), Maxwell's fisheye structure(of inhomogeneous refractive indices) manipulates light to transit smoothly from uniform space into curved space instead of just scattering the light with uniform permittivity. The different radiation patterns of two kinds of Maxwell fisheye medium, also inspires one to wonder other possible radiation patterns due to other possible but more complicated curved geometry of light.

However, solely from electron's circular trajectory on the lower half sphere in virtual uniform space for impedance-matched Maxwell fisheye profile, we are able to see how electrons perceive this curvature of EM space thus radiate transition radiation. Therefore the radiation from otherwise curved space will bend the electron's trajectory in uniform space to be along other curves (instead of spherical circles), and radiation pattern could contain even more side-lobes. This observation inspires one to engineer certain radiation patterns by varying permittivity and permeability in space and even add in anisotropicity similar to engineering directional beam in \cite{Leonhardt2008, Garcia2011}.  

\emph{Remark:} In this manuscript material dispersion is not considered for the simplicity of Green function because the emphasis is put on the spatial profile of refractive index which dominantly shapes the electromagnetic wave. Any imposed dispersion relation of $\epsilon(r, \omega)$ and $\mu(r,\omega)$ could further complicate the derivation of the relevant Green function. Thus this calculation is only valid within a narrow frequency band where our parameter profile applies.  

In conclusion, this manuscript investigates radiation from electron's motion through a curved space--Maxwell's fisheye profile. We calculate radiation from Maxwell's fisheye profile (both nonmagnetic and impedance-matched cases) using Green's function method, and finds that it combines both \u{C}erenkov radiation and transition radiation. For the impedance-matched Maxwell's fisheye, we also compare physical space and virtual uniform space to explain the reason electron's motion induces radiation in such a curved space geometry. Our calculation may point to novel methods to manufacture light source pumped from electron's kinetic energy. We also believe it is useful to explore possible radiation characteristics(radiation pattern) from the perspective of engineering permittivity and permeability. It is worth mentioning that electron-induced surface plasmon is also possible to occur when swift electrons interact with metal structure~\cite{Garcia2010}. This is out of scope of this manuscript, but also shares physics of transfer from evanescent wave to radiation. 

\section*{Methods}
In this section, we explain our methods to treat electromagnetic fields for radiation from the nonmagnetic Maxwell's fisheye sphere. This is a problem to solve inhomogeneous partial differential equations in mathematics. We first write Maxwell's equations in frequency domain according to Fourier transform. Second, we separate the whole current density or the source going through whole infinite space into two parts. The solution in uniform space(outside Maxwell's fisheye sphere), is also calculated through dyadic Green's function. Interested readers shall see more detail in Supplementary information I(iv). The other one inside Maxwell's fisheye sphere shall be solved via dyadic Green's function narrated in Supplementary information I(i-iii). The sum of both field solutions above make up the complete solution of electromagnetic fields. After that, inverse Fourier transform provides us solutions in time domain:
\begin{equation}
\bar E(\bar r,t)=\int_{-\infty}^{\infty}{\rm d}\omega\bar E(\bar r;\omega)e^{-i\omega t},
\end{equation}
Famous Fast-Fourier Transform can be a numerical solution to this infinite transform~\cite{Bracewell2000, Bailey1993}. The explicit expression of Green tensor for the full Maxwell fisheye in impedance-matched case is also given in Supplementary information III.

 \section*{Acknowledgments}
This work was partially supported by Singapore-MOE-AcRF 2008-T2-01-033. 
The computing resources are supported by the High Performance Computing (HPC) Cluster, Nanyang Technological University. 
L. K. Ang would like to acknowledge the support of SUTD start up grant (SERP11014 and IDSF 1200102).
Y. Liu would like to thank Leong Hon Wai, Zhang Junbin for their persistent assistance on configuration setup for parallel-computing, and Drs. Zhang Baile, Scott Robertson and Prof. U. Leonhardt for their helpful suggestions to improve this manuscript.

\section*{Author contributions}
YL conceived this idea of calculating radiation from Maxwell's fisheye, performed the calculation and wrote the manuscript. LKA supervised the project and revised the manuscript. 

\section*{Additional information}
Competing financial interests: The authors declare no competing financial interests.

Supplementary information accompanies this paper to report the derivation to treat EM fields for the radiation from Maxwell's fisheye. A gif file is also attached to display electric field patterns with time for impedance-matched Maxwell fisheye medium.





%








\section*{Supplementary information: Motion-induced radiation from electrons moving in Maxwell's fish-eye}


\section{Dyadic Green function method to solve electromagnetic fields in inhomogeneous medium}
In this appendix, we shall derive dyadic Green's function for Maxwell's fisheye sphere(nonmagnetic case), based on Tai's primer book~\cite{Tai1993}. 
For a source problem for electromagnetic fields, two \emph{inhomogeneous} wave equations below require solving,
\begin{eqnarray}\label{1}
\nabla\times\nabla\times\bar{E}-k^2\epsilon_r(r)\bar{E}&=&i\omega\mu_0\bar{J}\\
\label{2}
\nabla\times[\frac{1}{\epsilon_r(r)}\nabla\times\bar{H}]-k^2\bar{H}&=&\nabla\times\frac{\bar J}{\epsilon_r(\bar{r})}.
\end{eqnarray}
Main steps of dyadic Green's function to solve electromagnetic field with source are as follows, 

(i) Define four inhomogeneous spherical vector eigenwaves ${\bar M}(\bar r)$ and ${\bar N}(\bar r)$ from two homogeneous wave equations without source below: 
\begin{eqnarray}\label{3}
\nabla^2\Psi+k^2\epsilon_r(r)\Psi&=&0\\
\label{4}
\nabla^2\Phi-\frac{1}{\epsilon_r(r)}\frac{\mathrm{d}\epsilon_r(r)}{\mathrm{d}r}\frac{\partial \Phi}{\partial r}+k^2\epsilon_r(r)\Phi&=&0.
\end{eqnarray}
For any defined source function
\begin{eqnarray}\label{SJrt}
{\bar J}({\bar r},t)={\hat z}qv\delta(x-x_0)\delta(y-y_0)\frac{1}{\sigma\sqrt{2\pi}}\exp\Big[{-\frac{(z-vt)^2}{2\sigma^2}}\Big],
\end{eqnarray}
 four spherical vector eigen-waves,
\begin{eqnarray}
\label{veceig1}\bar{M}^{(m)}&=&\nabla\times(\Psi\bar{R})\\
\label{veceig2}\bar{N}^{(m)}&=&\frac{1}{k}\nabla\times\nabla\times(\Psi\bar{R})\\
\label{veceig3}\bar{M}^{(e)}&=&\nabla\times(\Phi\bar{R})\\
\label{veceig4}\bar{N}^{(e)}&=&\frac{1}{k\epsilon_r(r)}\nabla\times\nabla\times(\Phi\bar{R}),
\end{eqnarray}
are induced to solve (\ref{1}) and~\eqref{2}, providing that the generating functions, scalar quantities $\Psi$ and $\Phi$ satisfy following two differential equations \eqref{3} and \eqref{4}. Here piloting vector is defined as $\bar R=r\hat r$.

The symmetrical relations of these four vector eigenwaves are
\begin{eqnarray}
\label{veceig12}\bar{M}^{(m)}=\frac{1}{k\epsilon_r(r)}\nabla\times\bar{N}^{(m)}\\
\label{veceig22}\bar{N}^{(m)}=\frac{1}{k}\nabla\times\bar{M}^{(m)}\\
\label{veceig32}\bar{M}^{(e)}=\frac{1}{k}\nabla\times\bar{N}^{(e)}\\
\label{veceig42}\bar{N}^{(e)}=\frac{1}{k\epsilon_r(r)}\nabla\times\bar{M}^{(e)}
\end{eqnarray}
Note that both $\bar{M}^{(m)}$ and $\bar{N}^{(e)}$ are solutions for homogeneous vector wave equation of electric field $\bar{E}$, 
\begin{equation}
\nabla\times\nabla\bar{E}-k^2\epsilon_r(r)\bar{E}=0,
\end{equation}
while  both $\bar{M}^{(e)}$ and $\bar{N}^{(m)}$ are solutions for vector wave equation of magnetic field $\bar{H}$, 
\begin{equation}
\nabla\times[\frac{1}{\epsilon_r(r)}\nabla\times\bar{H}]-k^2\bar{H}=0.
\end{equation}
Thus with traditional vector spherical harmonics method in convenient spherical coordinates, by method of separation of radial and transverse components, scalar eigenwaves can be written explicitly
\begin{eqnarray}
\label{Psi}\Psi_n&=&\frac{1}{r}S_n(kr)P_n^m(\cos \theta)\exp(im\phi)\\
\label{Phi}\Phi_ n&=&\frac{1}{r}T_n(kr)P_n^m(\cos \theta)\exp(im\phi),
\end{eqnarray}
where radial components $S_n(kr)$ and $T_n(kr)$ satisfy ordinary differential equations
\begin{eqnarray}
\label{dSnr}
\frac{\mathrm{d}^2 S_n}{\mathrm{d}r^2}+[k^2\epsilon_r (r)-\frac{n(n+1)}{r^2}]S_n=0,\\
\label{dTnr}
\epsilon_r (r)\frac{\mathrm{d}}{\mathrm{d}r}[\frac{1}{\epsilon_r (r)}\frac{\mathrm{d}T_n}{\mathrm{d}r}]+[k^2\epsilon_r (r)-\frac{n(n+1)}{r^2}]T_n=0,
\end{eqnarray}
according to Eqs.~\eqref{3} and \eqref{4}.

(ii) Let us now denote the interior Maxwell's fisheye as region 1, and the exterior as region 2. The source coordinate of the dyadic Green's function is in region 1. Linear combinations of vector eigenfunctions ${\bar M}$ and ${\bar N}$  are juxtaposed into dyadic Green's function ${\bar{\bar G}_e}(\bar r, \bar r')$ in which the two labels in superscripts of Green's functions denotes the region in which the image coordinate is situated. In this part (ii), we only consider Green function when the source lies in region 1, so number 1 always appears as the second label herein. Now we have
\begin{equation}\label{Ge11}
\bar{\bar{G}}_e^{(11)}=\bar{\bar{G}}_{eo}(\bar{r},\bar{r}')+\bar{\bar{G}}_{es}^{(11)}(\bar{r},\bar{r}'), r<R_1
\end{equation}
and
\begin{equation}\label{Ge21}
\bar{\bar{G}}_e^{(21)}=\bar{\bar{G}}_{es}^{(21)}(\bar{r},\bar{r}'), r>R_1.
\end{equation}
\begin{eqnarray}\label{Geo}
\bar{\bar{G}}_{eo}&=&-\frac{1}{k^2}\hat{r}\hat{r}\delta(\bar{r}-\bar{r}')+\frac{ik}{4\pi}\sum_{m,n}C_{mn}\begin{cases}
\bar{M}^{(m1)}(k) \bar{M}'^{(m)}(k) + \bar{N}^{(e1)}(k)\bar{N}'^{(e)}(k), & r>r',\\
\bar{M}^{(m)}(k) \bar{M}'^{(m1)}(k) + \bar{N}^{(e)}(k)\bar{N}'^{(e1)}(k),& r<r'.\\
\end{cases}\\
\bar{\bar{G}}_e^{(11)}&=&\frac{ik}{4\pi}\sum_{n=1}^{\infty}\sum_{m=0}^{n}C_{mn}[A_n\bar{M}^{(m)}(k)\bar{M}'^{(m)}(k)+B_n\bar{N}^{(e)}(k)\bar{N}'^{(e)}(k)],\\
\bar{\bar{G}}_e^{(21)}&=&\frac{ik}{4\pi}\sum_{n=1}^{\infty}\sum_{m=0}^{n}C_{mn}[C_n\bar{M}^{(1)}(k)\bar{M}'^{(m)}(k)+D_n\bar{N}^{(1)}(k)\bar{N}'^{(e)}(k)].
\end{eqnarray}
\begin{equation}\label {defCmn}
C_{mn}=\frac{2n+1}{n(n+1)}(-1)^m. 
\end{equation}
Notice that for every primed vector in these dyads, label $m$ has to be replaced with $-m$. This coefficient is different from Tai's definition because we use $e^{\pm im\phi}$ instead of $\binom{\cos m\phi}{\sin m\phi}$. The dyadic Green functions above are written in analogy with their counterparts in uniform space. The dyadic Green's function for uniform space write(c.f. pp.198--203, \cite{Tai1993}),  
\begin{eqnarray}
\bar{\bar{G}}_{eo}&=&-\frac{1}{k^2}\hat{r}\hat{r}\delta(\bar{r}-\bar{r}')+\frac{ik}{4\pi}\sum_{m,n}C_{mn}\begin{cases}
\bar{M}^{(1)}(k) \bar{M}'^{}(k) + \bar{N}^{(1)}(k)\bar{N}'^{}(k), & r>r',\\
\bar{M}^{}(k) \bar{M}'^{(1)}(k) + \bar{N}^{}(k)\bar{N}'^{(1)}(k),& r<r'.\\
\end{cases}\\
\bar{\bar{G}}_e^{(11)}&=&\frac{ik}{4\pi}\sum_{n=1}^{\infty}\sum_{m=0}^{n}C_{mn}[A_n\bar{M}^{}(k)\bar{M}'^{}(k)+B_n\bar{N}^{}(k)\bar{N}'^{}(k)],\\
\bar{\bar{G}}_e^{(21)}&=&\frac{ik}{4\pi}\sum_{n=1}^{\infty}\sum_{m=0}^{n}C_{mn}[C_n\bar{M}^{(1)}(k)\bar{M}'^{}(k)+D_n\bar{N}^{(1)}(k)\bar{N}'^{}(k)];
\end{eqnarray}
whose building blocks, the eigenwave functions are defined as below:

\begin{align}
\bar{M}^{}=\nabla\times[j_n(kr)P_n^m(r)e^{im\phi}\bar{R}]\\
\bar{N}^{}=\frac{1}{k}\nabla\times\bar{M},\\
\bar{M}^{(1)}=\nabla\times[h^{(1)}_n(kr)P_n^m(r)e^{im\phi}\bar{R}]\\
\bar{N}^{(1)}=\frac{1}k\nabla\times\bar{M}^{(1)}.
\end{align}

The derivation of Green function $\bar{\bar{G}}_e$ for electric field of uniform space shall be briefly repeated here for clarity. Once it is done, one can simply replace eigenvectors $\bar M, \bar N, \bar M^{(1)}, \bar N^{(1)}$  into $\bar{M}'^{(m)},  \bar{N}'^{(e)}, \bar{M}'^{(m1)},  \bar{N}'^{(e1)}$; because radius functions $S_n(kr)$ and $T_n(kr)$ play the same role as $rj_n(kr)$ in the uniform(homogeneous) medium case, thus naturally we have the replacement above for all the eigenvectors(c.f. p260, ~\cite{Tai1993}). 

Now we repeat the main steps to derive the electric Green function for uniform space. Usually the Ohm-Rayleigh method to derive the Green function $\bar{\bar{G}}_e$, starts from equation on magnetic Green function $\bar{\bar{G}}_m$ and its relation equation with the electric one,
\begin{align}
\nabla\times\nabla\times\bar{\bar{G}}_m-k^2\bar{\bar{G}}_m=\nabla\times\bar{\bar{I}}\delta(\bar r-\bar r'),\\
\label{relationme}
\nabla\times\bar{\bar{G}}_m=\bar{\bar{I}}\delta(\bar r-\bar r')+k^2\bar{\bar{G}}_e.
\end{align}
Making use of orthogonal properties of spherical vector wave functions
\begin{align}
\iiint{\rm d}V\bar M_{mn}(\kappa)\cdot\bar N_{-m'n'}(\kappa')&=&0&,\\
\iiint{\rm d}V\bar M_{mn}(\kappa)\cdot\bar M_{-m'n'}(\kappa')=\iiint{\rm d}V\bar N_{mn}(\kappa)\cdot\bar N_{-m'n'}(\kappa')&=&\frac{\pi^2n(n+1)}{\kappa^2(2n+1)}&(-1)^m\delta_{nn'}\delta_{mm'}\delta(\kappa-\kappa'),
\end{align}
one can obtain
\begin{align}
\nabla\times\delta(\bar r-\bar r')=\frac{1}{2\pi^2}\int_0^\infty{\rm d}\kappa\sum_{m,n}C_{mn}
\kappa^3[{\bar N}(\kappa){\bar M'}(\kappa)+\bar M(\kappa){\bar N'}(\kappa)].
\end{align}
We have replaced $k$ into $\kappa$ to avoid duplication in the equations above and $C_{mn}$ is defined in equation~\eqref{defCmn}. Accordingly, 
\begin{align}
\bar{\bar{G}}_m(\bar r,\bar r')=\frac{1}{2\pi^2}\int_0^\infty{\rm d}\kappa\sum_{m,n}\frac{C_{mn}\kappa^3}{\kappa^2-k^2}[{\bar N}(\kappa){\bar M'}(\kappa)+\bar M(\kappa){\bar N'}(\kappa)]. 
\end{align}
And Green function $\bar{\bar{G}}_e$ in equation~\eqref{Geo} can be obtained by the help of equation~\eqref{relationme}. Q.E.D.

For magnetic fields $\vec H$ of our nonmagnetic Maxwell fisheye case, 
\begin{equation}
\bar{\bar{G}}_m^{(11)}=\bar{\bar{G}}_{mo}(\bar{r},\bar{r}')+\bar{\bar{G}}_{ms}^{(11)}(\bar{r},\bar{r}'), r<R_1
\end{equation}
and
\begin{align}
\bar{\bar{G}}_m^{(21)}=\bar{\bar{G}}_{ms}^{(21)}(\bar{r},\bar{r}'), r>R_1.\\
\bar{\bar{G}}_{m}=\nabla\times\bar{\bar{G}}_{e}:\\
\bar{\bar{G}}_{mo}&=&\frac{ik^2}{4\pi}\sum_{m,n}C_{mn}\begin{cases}
\bar{N}^{(m1)}(k) \bar{M}'^{(m)}(k) + \bar{M}^{(e1)}(k)\bar{N}'^{(e)}(k), & r>r',\\
\bar{N}^{(m)}(k) \bar{M}'^{(m1)}(k) + \bar{M}^{(e)}(k)\bar{N}'^{(e1)}(k),& r<r'.
\end{cases}\\
\bar{\bar{G}}_{ms}^{(11)}&=&\frac{ik^2}{4\pi}\sum_{n=1}^{\infty}\sum_{m=0}^{n}C_{mn}[A_n\bar{N}^{(m)}(k)\bar{M}'^{(m)}(k)+B_n\bar{M}^{(e)}(k)\bar{N}'^{(e)}(k)],\\
\bar{\bar{G}}_{ms}^{(21)}&=&\frac{ik^2}{4\pi}\sum_{n=1}^{\infty}\sum_{m=0}^{n}C_{mn}[C_n\bar{N}^{(1)}(k)\bar{M}'^{(m)}(k)+D_n\bar{M}^{(1)}(k)\bar{N}'^{(e)}(k)].
\end{align}
According to method of scattering superposition, coefficients $A_n$, $B_n$, $C_n$ and $D_n$ are determined from boundary conditions at lens boundary $r=R_1$(take the first branch of $r=R_1\geq r'$):
\begin{eqnarray}
{\hat r}\times {\bar E}={\hat r}\times {\bar E}\Longrightarrow {\hat r}\times \bar{\bar{G}}_{e}^{(11)}&=&{\hat r}\times \bar{\bar{G}}_{e}^{(21)}\\
{\hat r}\times {\bar H}={\hat r}\times {\bar H},\nabla\times{\bar E}=i\omega\mu_0{\bar H} \Longrightarrow {\hat r}\times \nabla\times\bar{\bar{G}}_{e}^{(11)}&=&{\hat r}\times \nabla\times\bar{\bar{G}}_{e}^{(21)}.
\end{eqnarray}
Therefore a set of algebraic equations are deduced, 
\begin{eqnarray}
S^{(1)}+A_nS=C_nQ,\\
T^{(1)}+B_nT=D_nQ,\\
S'^{(1)}+A_nS'=C_nQ',\\
T'^{(1)}+B_nT'=D_nQ';
\end{eqnarray}
where
\begin{eqnarray}
S^{(1)}=kS_n^{(1)}(\rho_a), \\
S=kS_n(\rho_a),\\
Q=\rho_a h_n^{(1)}(\rho_a),\\
Q'=\frac{\rm d}{{\rm d}\rho_a}S_n(\rho_a),\\
T=kT_n(\rho_a),\\
T'=\frac{\rm d}{{\rm d}\rho_a}kT_n(\rho_a),\\
\rho_a=kr_1.
\end{eqnarray}
Once we solve the four coefficients , we know the dyadic Green functions when source is located in region 1. 

(iii) Multiplied by current source term $\bar{J}(\omega;\bar r)$, dyadic Green's function ${\bar{\bar G}_e}(\bar r, \bar r')$ gives electromagnetic field solution ${\bar E}(\omega; {\bar r})$ explicitly. For a line-section of current density ${\bar J}({\bar r},\omega)$ in \eqref{SJrt}, only the line section penetrating the Maxwell's lens inside (zone 1) is to be treated using the dyadic Green functions in Eqs.~\eqref{Ge11} and~\eqref{Ge21}, and we write
\begin{eqnarray}
\label{E3}
\bar{E}(\bar{r}) = i\omega\mu_0\iiint\bar{\bar{G}}_e(\bar{r}, \bar{r'})\cdot\bar {J}(\bar {r'})\mathrm d V',\\
\bar H(\bar r) = \iiint\nabla\times\Bar{\Bar G}_e(\bar r, \bar r')\cdot\bar J(\bar r')\mathrm d V'.
\end{eqnarray}
Let us write current density $\bar J(\bar r,t)$ in Eq.~\eqref{SJrt} as $\bar J_z={\hat z}J_3$ into spherical coordinates(curvilinear orthogonal coordinates~\cite{Leonhardt2009b,Tai1992,Tai1993}). The identity to be used here writes ${\rm d}r={\rm d}x^i{\hat e_i}={\rm d}x^j{\hat e_j}=\frac{\partial x^{i'}}{\partial x^j}{\rm d}x^j{\hat e_{i'}}\Rightarrow \frac{\partial x^{i'}}{\partial x^j}{\hat e_{i'}}={\hat e_{j}}$, so $\hat {z'}=\hat{r'}\cos\theta'-\hat{\theta'}\sin\theta'$. To further implement line integration along z coordinate, integrand above
\begin{eqnarray}\label{}
\Bar{\Bar G}_e(\bar r, \bar r')\cdot\bar J(\bar r')
=\Bar{\Bar G}_e(\bar r, \bar r')\cdot{\hat z'}J_3=\Bar{\Bar G}_e(\bar r, \bar r')\cdot (\hat r' \cos\theta'-\hat\theta'\sin\theta') J_3,
\end{eqnarray}
which finally has to be converted into Cartesian coordinates to implement the line integral.  In such a step, a vector in spherical coordinates $(r,\theta,\phi)$converts into one in Cartesian coordinates $(x,y,z)$ according to
\begin{eqnarray}\label{}
x&=&r\sin\theta\cos\phi,\\
y&=&r\sin\theta\sin\phi,\\
z&=&r\cos\theta.
\end{eqnarray}

(iv) The remaining infinite long line of current density in uniform space, we shall resort to traditional dyadic Green's method. Current density in frequency domain
\begin{equation}
{\bar J}({\bar r},\omega)={\hat z}\frac{q}{2\pi}\delta(x-x_0)\delta(y-y_0)\exp\big({-\frac{\sigma^2\omega^2}{2v^2}+i\frac{\omega z}{v}}\big), 
\end{equation}
can be written in form of 
\begin{eqnarray}\label{Jpiece}
{\bar J}({\bar r},\omega)=\Big\{ {\bar J}({\bar r},\omega) -{\bar J}({\bar r},\omega)[u(z-z_1)-u(z-z_2)]\Big\}+{\bar J}({\bar r},\omega)[u(z-z_1)-u(z-z_2)]=\bar J_1+\bar J_2,
\end{eqnarray}
where 
\begin{eqnarray}
\bar J_1={\bar J}({\bar r},\omega) -{\bar J}({\bar r},\omega)[u(z-z_1)-u(z-z_2)],\\
\bar J_2={\bar J}({\bar r},\omega)[u(z-z_1)-u(z-z_2)],
\end{eqnarray}
in which $u(\cdot)$ represents step function and $z_1, z_2$ indicates $z$ coordinates of points where injection current line intersects Maxwell's fisheye. The terms $\bar J_1$  and $\bar J_2$ is respectively positioned in the interior and exterior part of Maxwell's fisheye sphere,  therefore they should be treated with different dyadic Green's functions due to different EM space structures.  According to Zhang's work~\cite{Zhang2009} and its supplementary material, the line-section $\bar J_1$ outside of Maxwell's fisheye can be considered as a difference between an infinite current line and a definite one, both of which can be treated straightforwardly~\cite{Kong_EWT}. Electric field solution for the infinite current line can be written in form of 
\begin{eqnarray}
\bar{E}_1(\bar{r},\omega)=-\frac{q\exp{\big({-\frac{\sigma^2\omega^2}{2v^2}+i\frac{\omega z}{v}}\big)}}{8\pi\omega\epsilon_0}\big[\hat z\Big(\frac{\omega^2}{c^2}-\frac{\omega^2}{v^2}\Big)H_0^{(1)}(k_\rho \rho)-\hat\rho\frac{i\omega k_\rho}{v}H_1^{(1)}(k_\rho \rho)\big],
\end{eqnarray}
in which
\begin{equation}
k_\rho=\sqrt{k^2-\frac{\omega^2}{v^2}},
\end{equation}
and
$H_i^{(1)}(\cdot)$ means $i$th order Hankel function of the first kind. For $\bar{E}_1(\bar{r},\omega)$above, the cylindrical system's origin is set as point $(x_0,y_0,0)$. 

Electric field solution for definite line segment inside uniform sphere $R_1$ can be obtained from equation
\begin{equation}
\bar{E_2}(\bar{r},\omega) = i\omega\mu_0\big[\bar{\bar{I}}+\frac{1}{k^2}\nabla\nabla\big]\cdot\iiint\frac{e^{ik\vert \bar r-\bar r'\vert}}{4\pi\vert\bar r-\bar r'\vert}  {\bar J}({\bar r'},\omega)[u(z'-z_1)-u(z'-z_2)] \mathrm d V'.
\end{equation}
Therefore the electric field we seek is, 
\begin{equation}
\bar{E}(\bar{r},\omega) = \bar{E_1}(\bar{r},\omega) -\bar{E_2}(\bar{r},\omega) +\bar{E_3}(\bar{r},\omega),
\end{equation}
in which $\bar{E_3}(\bar{r},\omega)$ is calculated from Eq.~\eqref{E3} in (iii) we have derived.

(v) It is straightforward to write radiation energy spectrum density for radiation from Maxwell's fisheye. The enclosing sphere used for calculation is defined as $r=2R_1=4{\rm \mu m}$. We use $\Re$ and $\Im$ to denote real and imaginary parts respectively. 
\begin{eqnarray}\label{dWdomega}
\frac{d W}{d \omega}&=&\int_0^{2\pi} \int_0^{\pi} r^2 \sin \theta
d\theta d\phi \cdot 4\pi \left[\Re\bar E(\omega) \times \Re
\bar H(\omega)  + \Im \bar E(\omega)  \times \Im \bar H(\omega) 
\right] \cdot \hat r.
\end{eqnarray}
The radiation pattern formula is the integrand of the integral above in Eq.\eqref{dWdomega}. 

\section{Wave solution in Maxwell's fisheye sphere}
This appendix reports how we solve scalar Helmholtz Eqs.~\eqref{dSnr} and \eqref{dTnr}.
For the Maxwell's fisheye lens, given permittivity function defined as
\begin{eqnarray}
\epsilon_r(r)= \frac{4}{\big[1+(\frac{r}{R_1})^2\big]^2} , r \in [0, R_1], 
\end{eqnarray}

Radius $r$ ranges from 0 to $R_1$. Tai transformed  radial variable in this way~\cite{Tai1958, Tai1993}, 
\begin{eqnarray}
\rho=kr, \rho_a=kR_1, \xi=-(\frac{\rho}{\rho_a})^2\\
S_n({\xi})=\xi^{\frac{n+1}{2}}(\xi-1)^\mu U_n(\xi)\\
T_n({\xi})=\xi^{\frac{n+1}{2}}(\xi-1)^{\mu-1} V_n(\xi),
\end{eqnarray}
where 
\begin{eqnarray}\label{}
\mu=\frac{1}{2}(1+\sqrt{1+4\rho_a^2}),
\end{eqnarray}
thus new functions $U_n(\xi)$ and $V_n(\xi)$ form two hypergeometric equations,
\begin{eqnarray}\label{5}
\xi(\xi-1)\frac{\mathrm{d}^2 U_n}{\mathrm{d}\xi^2}+[(2\mu+\beta)\xi-\beta]\frac{\mathrm{d} U_n}{\mathrm{d}\xi} + \alpha U_n&=& 0\\
\label{6}
\xi(\xi-1)\frac{\mathrm{d}^2 V_n}{\mathrm{d}\xi^2}+[(2\mu+\beta)\xi-\beta]\frac{\mathrm{d} V_n}{\mathrm{d}\xi} +(\alpha-\frac{1}{2}) V_n &=& 0,
\end{eqnarray}
where 
\begin{eqnarray}
\beta= n+\frac{3}{2},\\
\alpha = \beta \mu +\rho_a^2,
\end{eqnarray}
whose solutions are hypergeometric function $_2 F_1(a, b; c; \xi)$ and $\xi^{1-c}_2 F_1(1+a-c, 1+b-c; 2-c; \xi)$. 
Comparing relevant coefficients of Eqs. \eqref{5} and \eqref{6} with those of normal form of hypergeometric equation, 
\begin{eqnarray}\label{}
a+b+1=2\mu+\beta\\
c=\beta\\
ab=\alpha &or& \alpha -\frac{1}2.
\end{eqnarray}
Hence solutions to Eqs. \eqref{5} and \eqref{6} read, regular solution
\begin{eqnarray}\label{hyper1}
_2 F_1(a, b; c; z)\triangleq F(a,b,c,z)=F(a(2\mu+\beta, \alpha), b(2\mu+\beta, \alpha),\beta, -(\frac{\rho}{\rho_a})^2). 
\end{eqnarray}
and the solution singular at origin
\begin{eqnarray}\label{hyper2}
\xi^{1-c}_2 F_1(1+a-c, 1+b-c; 2-c; \xi)\triangleq \xi^{1-c}F^{(1)}(1+a-c, 1+b-c; 2-c; \xi), 
\end{eqnarray}
so we expect to determine four inhomogeneous spherical vector eigenwaves, $\bar{M}^{(m)}, 
\bar{N}^{(m)}, \bar{M}^{(e)} $, $\bar{N}^{(e)}$ and $\bar{M}^{(m1)}, 
\bar{N}^{(m1)}, \bar{M}^{(e1)} $, $\bar{N}^{(e1)}$ from Eqs.~\eqref{veceig1}, \eqref{veceig2}, \eqref{veceig3}, \eqref{veceig4};\eqref{Psi}, \eqref{Phi} and \eqref{hyper1}, \eqref{hyper2}. For instance, 
\begin{equation}
\bar{M}^{(m)}=\nabla\times(\Psi\bar{R}) , \Psi=\frac{1}{r}S_n(kr), S_n(kr)=F\Big(a(2\mu+\beta, \alpha), b(2\mu+\beta, \alpha),\beta, -\Big(\frac{r}{R_1}\Big)^2\Big).
\end{equation}
The rest form of vector eigenwaves are similar. 
 
Then dyadic Green's function $\Bar{\Bar G}_e(\bar r, \bar r')$ in Maxwell's fisheye sphere is therefore determined, which determines electric fields in the fisheye sphere according to equations in Section I (i-iii).

\section{Explicit form of Green tensor for the full impedance-matched Maxwell fisheye}

(i) Before writing out its explicit form of Green tensor, the electric field pattern from a Hertzian electric dipole inside the full impedance-matched Maxwell fisheye is plotted in Fig.~S1(a) to compare with that of uniform space in Fig.~S1(b). We adopt the same parameters as Fig.~2(a) in the maintext and observe a similar continuously-varying behavior of electric field.

\begin{figure}[htbp]
  \centering
	\includegraphics[width=1\columnwidth]{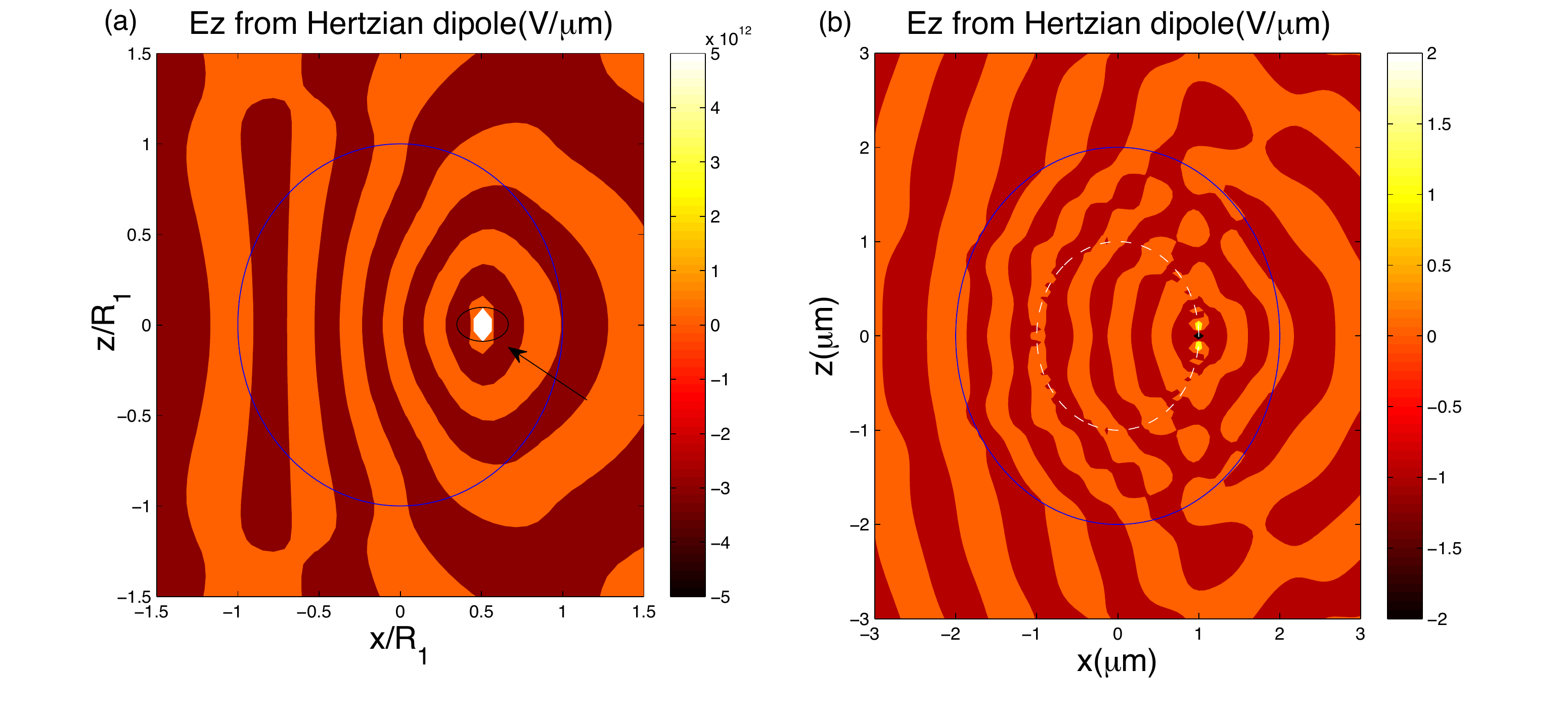}
\caption
{\label{fig2add} -S
(Color online)Electric fields in z direction, $E_z(x,z)$ on xz plane from Hertzian dipole at point $x=R_1/2=1\mu m, y=0, z=0$(marked as a small black circle) for (a) the full Maxwell's fisheye sphere in impedance-match case and (b)(Replica of Figure 2(b) in maintext) uniform sphere ($\epsilon_r(r<R_1)=4$).  
}
\end{figure}

(ii) The Green tensor for electromagnetic fields in the full Maxwell fisheye has been solved recently\cite{Leonhardt2010b, Leonhardt2011}, in which the Green tensor for electric field should be, 
\begin{eqnarray}\label{GtensorUlf}
G(\bar r,\bar r', k)&=&\frac{\nabla\times n(\gamma)\nabla\otimes\nabla'D(\gamma)\times \overleftarrow \nabla'}{n(r)n(r')k^2}-\frac{\delta(\bar r-\bar r') \mathbb {1}  }{n(r)k^2},\\
n(r)&=&\frac{2}{1+r^2},\\
{D}(\gamma,\omega)&=&\Big(\gamma+\frac{1}{\gamma}\Big)\frac{\sin(2\omega \;arccot\; \gamma)}{(4\pi)^2\sin \pi\omega},\\
\gamma&=&\frac{\vert\vec r-\vec {r'}\vert}{\sqrt{1+2\vec r\cdot \vec {r}' + \vert \vec r\vert^2\vert\vec r'\vert ^2 }}. 
\end{eqnarray}
in which the length quantities $r$ and $r'$ are taken in the relative unit of inner sphere radius $R_1$ and light speed $c=1$ to avoid unnecessary factors, however one requires to convert the unit after obtaining the physical quantities. Thus electric field in in the full Maxwell's fisheye sphere can be written the same as equation ~\eqref{E3}. The only non-trivial part for this Green tensor is the nominator of its first term, let us call it $K$. If we define($x,y,z$ are written as $x_1,x_2, x_3$ to facilitate tensor symbol)
\begin{equation}
D_{ij}=\partial_{x_i}\partial_{x'_j}D(r,r')(i=1,2,3), 
\end{equation}
\begin{eqnarray}\label{}
K(\bar r,\bar r',k)&=&\nabla\times n(r,r')\nabla\otimes\nabla' D(r,r')\times \overleftarrow \nabla'\\
&=&\nabla\times n(r,r')D_{ij}\hat {x_i}\hat{x}'_{j}\times \overleftarrow \nabla'\\
&=&\nabla\times n(r,r')\begin{pmatrix} D_{11} & D_{12} & D_{13}\\ D_{21} & D_{22} & D_{23}\\D_{31} & D_{32} & D_{33}\end{pmatrix}\times \overleftarrow \nabla'\\
&=&[acd][bef]\frac{\partial^2n(r,r')}{\partial x_c \partial x'_e}\frac{\partial^2D(r,r')}{\partial x_d\partial x'_f}\hat {x}_a\otimes\hat{x'}_{b},\\
K\cdot\hat{z'}&=&
\begin{bmatrix} D_{32} \partial_{x'}\partial _y n(r,r')- D_{22} \partial_{x'}\partial_z n(r,r')\\  
D_{12} \partial_{x'}\partial _z n(r,r')- D_{32} \partial_{x'}\partial_x n(r,r')\\
D_{22} \partial_{x'}\partial _x n(r,r')- D_{12} \partial_{x'}\partial_y n(r,r')\end{bmatrix}-
\begin{bmatrix} D_{31} \partial_{y'}\partial _y n(r,r')- D_{21} \partial_{y'}\partial_z n(r,r')\\  
D_{11} \partial_{y'}\partial _z n(r,r')- D_{31} \partial_{y'}\partial_x n(r,r')\\
D_{21} \partial_{y'}\partial _x n(r,r')- D_{11} \partial_{y'}\partial_y n(r,r')\end{bmatrix}.
\end{eqnarray}
To arrive at this step, the rule that a curl of a gradient vanishes is used, as well as a little tensor algebra~\cite{LebedevTensor}. Once we know how to write the explicit Green tensor, electric field is known according to equation~\eqref{E3}.  

Although from the temporal perspective, it is also possible write time-dependent Green tensor\cite{Leonhardt2011} in this case, however, to obtain electric field solution, explicit derivation shows that the only trouble is to solve a transcend equation, which happens to be much more messy than in frequency domain. That is why we adopt Green tensor in frequency domain.  

(iii) A gif file is also attached to display electric field patterns with time for impedance-matched Maxwell fisheye medium.

\bibliographystyle{naturemag}
\bibliography{bib5}

\end{document}